\documentclass[useAMS,usenatbib]{mn2e}

\usepackage{graphicx}

\RequirePackage{xspace}
\RequirePackage{amsmath}
\RequirePackage{amsfonts}
\RequirePackage{amssymb}

\newcommand{\phx}{\texttt{PHOENIX}\xspace}

\title[Time series of SN~2014J spectra with TIGRE]
{Time series of high resolution spectra of SN~2014J observed 
with the TIGRE telescope}
\author[D. Jack et al.]{D. Jack,$^{1}$\thanks{E-mail:
dennis@astro.ugto.mx} M. Mittag,$^{2}$ K.-P. Schr\"oder,$^{1,2}$
J. H. M. M. Schmitt,$^{2}$ A. Hempelmann,$^{2}$\newauthor 
J. N. Gonz\'alez-P\'erez,$^{2}$ M. A. Trinidad,$^{1}$ G. Rauw$^{3}$ and 
J. M. Cabrera Sixto$^{4}$\\
$^{1}$Departamento de Astronom\'\i{}a, Universidad de Guanajuato, A.P.~144, 
      36000 Guanajuato, GTO, Mexico\\
$^{2}$Hamburger Sternwarte, University of Hamburg, Gojenbergsweg 112, 
      21029 Hamburg, Germany\\
$^{3}$Groupe d'Astrophysique des Hautes Energies, Institut 
d'Astrophysique et de 
G{\'e}ophysique, Universit{\'e} de Li{\`e}ge, All{\'e}e du 6 Ao{\^ut},\\
B{\^a}t B5c, 4000 Li{\`e}ge, Belgium\\
$^{4}$Universidad de Guanajuato, Lascur\'ain de Retana No 5, C.P. 
      36000 Guanajuato, GTO, Mexico}
      
\begin{document}

\date{Accepted xxx Received xx; in original form xxx}

\pagerange{\pageref{firstpage}--\pageref{lastpage}} \pubyear{2015}

\maketitle

\label{firstpage}

\begin{abstract}
We present a time series of high resolution spectra of the Type Ia supernova 2014J,
which exploded in the nearby galaxy M82.
The spectra were obtained with the HEROS \'echelle spectrograph installed at
the 1.2~m TIGRE telescope.
We present a series of 33 spectra with a resolution of $R\approx 20,000$,
which covers the important bright phases in the evolution of SN~2014J
during the period from January 24 to April 1 of 2014.
The spectral evolution of SN 2014J is derived empirically. The expansion velocities
of the Si~II P-Cygni features were measured and show the expected decreasing behaviour,
beginning with a high velocity of 14,000~km/s on January 24.
 The Ca~II infrared triplet feature shows a high velocity component with
expansion velocities of $>20,000~{\rm km/s}$ during the early evolution apart
from the normal component showing similar velocities as Si~II.
Further broad P-Cygni profiles are exhibited by the principal 
lines of Ca~II, Mg~II and Fe~II. The TIGRE SN~2014J spectra also resolve several very sharp 
Na~I D doublet absorption components. Our analysis suggests interesting substructures 
in the interstellar medium of the host galaxy M82, as well as in our Milky Way, 
confirming other work on this SN. 
We were able to identify the interstellar absorption of M82 in the lines 
of Ca~II H \& K at 3933 and 3968~\AA\ as well as K~I at 7664 and 7698~\AA.
Furthermore, we confirm several 
Diffuse Interstellar Bands, at wavelengths of 
6196, 6283, 6376, 6379 and 6613~\AA\ and give their measured equivalent widths.

\end{abstract}


\section{Introduction}

Supernovae of Type Ia (SN Ia) are of special interest for cosmology since the
discovery that the expansion of the universe is actually accelerated 
\citep{riess_scoop98,perletal99}.
But despite being widely used as a calibratable ``standard candle''
\citep{philm15,rpk96,philetal99,goldhetal01} this type of supernovae remains
poorly understood. See \citet{parrent14} for a recent review on SNe Ia and 
their properties.
The physical nature of a Type Ia supernova progenitor is still under discussion
\citep{levanon15} as well as the explosion mechanism. Several
different explosion models have been suggested and calculated
\citep{nomoto84,khok91a,PCL04,jordangcd08,pakmor12,pakmor13,rosswog09,kushnir13}.
A well-resolved time series 
of spectra of SNe Ia may shed some light on the physical properties of the 
expanding envelope and, therefore, might reveal some vital clues about the
nature of SN Ia explosions.

So far, however, no such detailed time series of SNe Ia spectra exist.
The most detailed time series obtained of SN 2011fe has only a relatively low
spectral resolution and some gaps in the coverage \citep{pereira13}.
The nearby and, therefore, quite bright
supernova 2014J provided an excellent opportunity for observing a SN~Ia very
closely and obtaining quite high resolution spectra.

SN~2014J was discovered in the starburst 
galaxy M82 \citep{fossey14} during the night of January 21, in an early 
stage of its outbrake. Only two weeks later, around February 4, the 
maximum $V$-magnitude was reached with $V = 10.5$. By mid-march, SN 2014J 
had faded below 12.0 mag, but it was still well placed in the sky 
for good-quality observations. It has been observed by many telescopes and
in different wavelength regions \citep{telesco14,foley14b,ashall14,
margutti14,perez14}, however, none of them have obtained
observations with a very detailed time series of optical spectra with
higher resolution.

As an additional bonus of this bright supernova, we can use it simply as a
point light source to probe the interstellar and perhaps intergalactic
medium in the line of sight, mostly belonging to the supernova host
galaxy and the Milky Way.
Therefore, high resolution spectra of supernova explosions in other galaxies
reveal also information about substructures in
the interstellar medium (ISM) in our galaxy and in the
respective host galaxy.
The lines of interest are those of suitable ion
species of metals showing resonance lines (e.g. Na~I, Ca~II, Mg~II), which
represent transitions from the ground level, e.g. with a lower level of
zero excitation energy. In the optical, the sodium D line is most suitable
to probe interstellar matter and its dynamics. First high-resolution
spectra of this kind were obtained and interpreted on the extraordinary
opportunity presented by the bright SN 1987A \citep{deboer87}. Nearly 
30 years of advances in detector technology allow such
studies to be undertaken with much less bright supernovae and more modest
equipment.

Other still unidentified features of interstellar absorption, 
which can be observed in SN spectra,
are the Diffuse Interstellar Bands (DIB).
In our Galaxy, such DIBs are quite commonly observed in high-resolution spectra
of massive stars \citep{herbig95}. There
exist also measurements for the Small and Large Magellanic Clouds
\citep{vladilo87,ehrenfreund02,cox06,cox07,welty06}.
Some spectral observations
of extragalactic SNe have already succeeded in detecting DIBs \citep{dodorico89,
sollerman05,cox08,cox14}. For a further recent study of DIBs a galactic
nucleus has been used as the background light source \citep{ritchey15b}.
Hence, the nearby SN 2014J presents a further opportunity
to study DIBs in high spectral resolution and to expand on the
work by \citet{welty14}.

There already exist some work on ISM observations in SN 2014J spectra.
A detailed analysis of the interstellar medium of M82 is presented in \citet{ritchey15}.
They use six spectra and determine some abundances.
\citet{welty14} did a thorough analysis of DIBs using the same set of spectra.
\citet{graham14} present a series of very high resolution spectra ($R\approx 110,000$)
of SN 2014J. They reveal many substructures in the Na~I D line. 
Furthermore, they identify other interstellar absorption lines and DIBs.

In the following section \ref{sec:obs}, we first introduce
the instrumentation used, the 1.2~m TIGRE telescope and its HEROS 
spectrograph, then present our observations: a detailed time series 
of 33 high resolution spectra of the Type Ia supernova 2014J in M82,
covering the period from January 24 to April 1. We continue with an 
empirical study of the evolution of the prominent Si II line and other 
spectral features. In a further section \ref{sec:NaD},
we take advantage of the high resolution of the TIGRE/HEROS spectra and
study the multi-component interstellar absorption in the Na I D absorption 
lines and elsewhere in the SN 2014J spectra.

\section[]{Spectroscopic monitoring of SN~2014J}\label{sec:obs}

\subsection{Instrumentation: el TIGRE}
The TIGRE telescope is a fully automated telescope with an aperture of 1.2~m,
situated near the city of Guanajuato in Central Mexico. It is equipped with the 
HEROS \'echelle spectrograph, which has a resolution of $R\approx20,000$.
Spectra are recorded simultaneously in two 
channels, blue and red, covering the large wavelength range from 3800 to 8800 {\AA}
with just a small gap of 130~\AA\ around 5800 {\AA}. Like operations, the data 
reduction pipeline is also fully automatic.

Originally designed to monitor point-like objects 
down to about 10th magnitude with high-quality spectra, SN 2014J presented a 
challenge as much as an opportunity for this relatively small telescope. 
For a more detailed technical description of the TIGRE instrumentation 
and its capabilities, see \citet{schmitt14}.

\subsection{Time series of SN 2014J spectra}

\begin{figure*}
\begin{center}
\includegraphics[width=\textwidth]{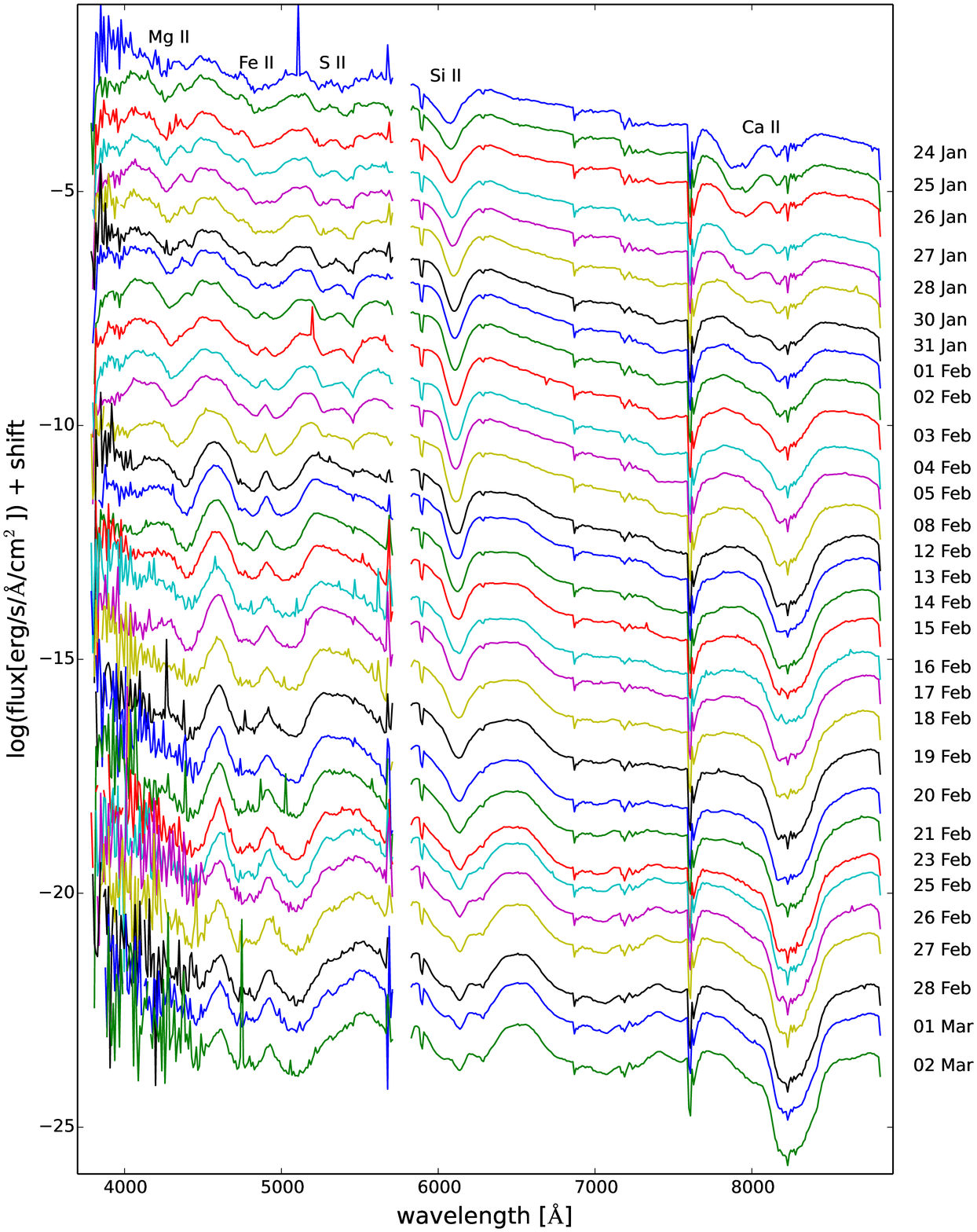}
\caption{Spectral time series of SN 2014J observations obtained with the TIGRE
telescope in both the red and blue channel of the HEROS spectrograph. The dates 
are given for the observations in early 2014.}
\label{fig:all_specs}
\end{center}
\end{figure*}

Our monitoring of the supernova 2014J started very shortly
after its discovery on January 21. We obtained high-resolution
spectra with a good signal to noise (S/N) of around 60 in the red channel
in almost every night until 
March 2. Due to some tecnical problems with the telescope we
were only able to take two late time spectra on March 31 and April 1.
In the blue channel the S/N
has lower values of around 20, because SN 2014J suffered from a significant
interstellar absorption and reddening. Therefore, 
all spectra presented in Fig. \ref{fig:all_specs} have been dereddened with
the values $E(B-V)=1.33$ and $R_{\rm V}=1.3$ found by 
\citet{amanullah14} in their study of the extinction law of SN 2014J.
Furthermore, since the full resolution is not necessary for a study of the broad 
spectral features of supernova ejecta, we binned our spectra to a resolution of 
$\Delta\lambda=10$~\AA\ for this purpose. Hence, the S/N was further improved and
the spectra smoothed, also reducing the visibility of some telluric lines
inherent to ground based observations.

Figure \ref{fig:all_specs} shows the complete set of SN 2014J spectra observed
with the TIGRE telescope.
We combined the observations of the blue and the red channel to present the full
spectra in the whole wavelength range from 3800 to 8800~\AA\ except
for the small gap of the spectrograph around 5800~\AA.
The dates at the right hand side mark the observation date in 2014 in Universal Time 
(UT). Because of the strong interstellar reddening of SN~2014J,
the later spectra in the blue channel have a poor S/N and one cannot
clearly distinguish features any more.

The very first spectrum was observed on January 24 UT and, therefore,
still about 10 days before maximum light.
This spectrum already shows the prominent P-Cygni profile of the
Si~II feature at around 6300~\AA, which usually
identifies a SN as a Type Ia. A high expansion velocity feature
of the Ca~II triplet is also present in the early spectra of SN~2014J
at a wavelength of around 8000~\AA.
We were able to observe a spectrum almost every night until March 2 UT.
In this time, 31 spectra were obtained. Then, due to a simple technical problem with 
the telescope mount, which caused some out-time for repair work, we could only 
obtain two more late-time spectra of SN 2014J, on March 31 and April 1 UT. 

A SN Ia spectrum at early times is supposed to be quite flat and should 
not show many clear features. This is clearly consistent with the
appearance of our early-time spectra of SN~2014J. For clarity, the latest 
spectrum shown in Fig.~\ref{fig:all_specs} is the one observed on March 2, which
corresponds to about 1 month after maximum light.
The small "lines" in some parts of the spectra stem from a few, 
more noticeable telluric lines. 

The here presented time series reveals the changing shape, on an almost daily 
schedule with some features. The general trend,
the P-Cygni extrema moving towards the redder part of the spectrum,
is consistently observed in the Si~II line and some other features. This 
means that the observed expansion velocities change during the evolution.
The reason for that is the expansion of the envelope, which
decreases the mass densities and, therefore, decreases all opacities. 
Hence, with ongoing expansion, ever deeper and slower layers of the expanding 
envelope become visible in the spectra. In other words, the quasi-photosphere 
moves inwards into less fast expanding shell material.

Another strong feature in SN Ia spectra is 
caused by the Ca~II IR triplet at around 8500~\AA\, which already appears during 
the early phases at around maximum light. It can be observed throughout the 
whole evolution until the latest spectrum shown in Fig.~\ref{fig:all_specs}.
At early times it shows a high velocity component.

At later phases an Fe II emission feature appears about where the prominent 
Si II line was observed earlier. Another Fe~II feature then appears at around 
5000~\AA.  Indeed, it is true of evolved type Ia SNe to show more 
spectral features in general. Eventually, these go into emission, as the late, 
optically thin phase of the expanding envelope has been reached. At the same time,
one sees the inner layers, which are rich of iron peak elements, causing the
above-mentioned rise of Fe~II features. In the following section, we present a
more detailed picture of the evolution of selected spectral features.

\subsection{Detailed spectral evolution of important features}

\begin{figure}
\begin{center}
 \resizebox{\hsize}{!}{\includegraphics{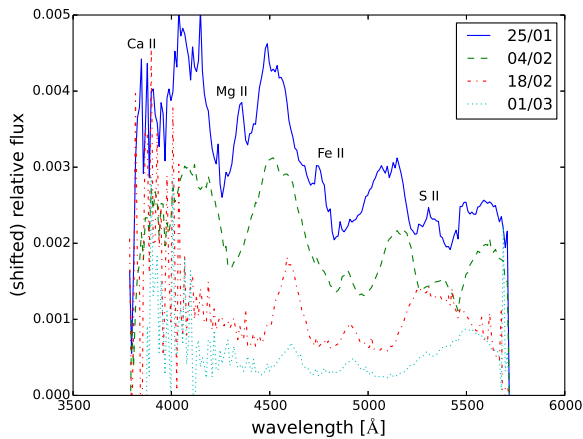}}
\caption{Evolution of selected SN 2014J spectra obtained in the blue channel of the
HEROS spectrograph.}
\label{fig:blue}
\end{center}
\end{figure}

In Fig.~\ref{fig:blue}, we compare a selection of only four spectra from the blue 
channel of the HEROS spectrograph, representative of four different epochs in the 
evolution of SN~2014J. Due to the strong interstellar reddening of the light of
SN~2014J, the later spectra have only poor S/N.

The first spectrum from above in Fig.~\ref{fig:blue} was obtained on January 25. 
It shows the typical features of a SN Ia spectrum during maximum light.
We can see the typical S II line profile in form of a W between wavelengths of, 
roughly, 5000 and 5500 \AA. At shorter wavelengths a clear Fe~II feature can be seen 
at around 4500 to 5000 \AA. Furthermore, around wavelengths of about 
4000 to 4500~\AA\, a clear feature of Mg~II is present in the spectrum of
SN~2014J. Below 4000~\AA\ one can see the drop in flux towards the Ca II H \& K
feature, although it is not fully covered in the shown spectrum and
the S/N is already quite low in that part of the spectrum. We should stress
that spectra of SNe Ia are a blend of millions of lines, so that one cannot
always assign one specific element to each observed feature.

The spectrum of February 4, which corresponds to maximum light, shows
similar spectral features, when compared to the spectrum of January 25.
However, the line profiles have already changed a bit. Additionally,
this spectrum shows consistently lower expansion velocities, in all of its 
features. The further two spectra shown in Fig.~\ref{fig:blue}, of February 
18 and the one of March 1, suffer already from poorer S/N. Nevertheless, 
these spectra show the expected emission features of Fe~II.

\subsubsection{Evolution of the Si II line profile}

\begin{figure*}
\begin{center}
\includegraphics[scale=0.75]{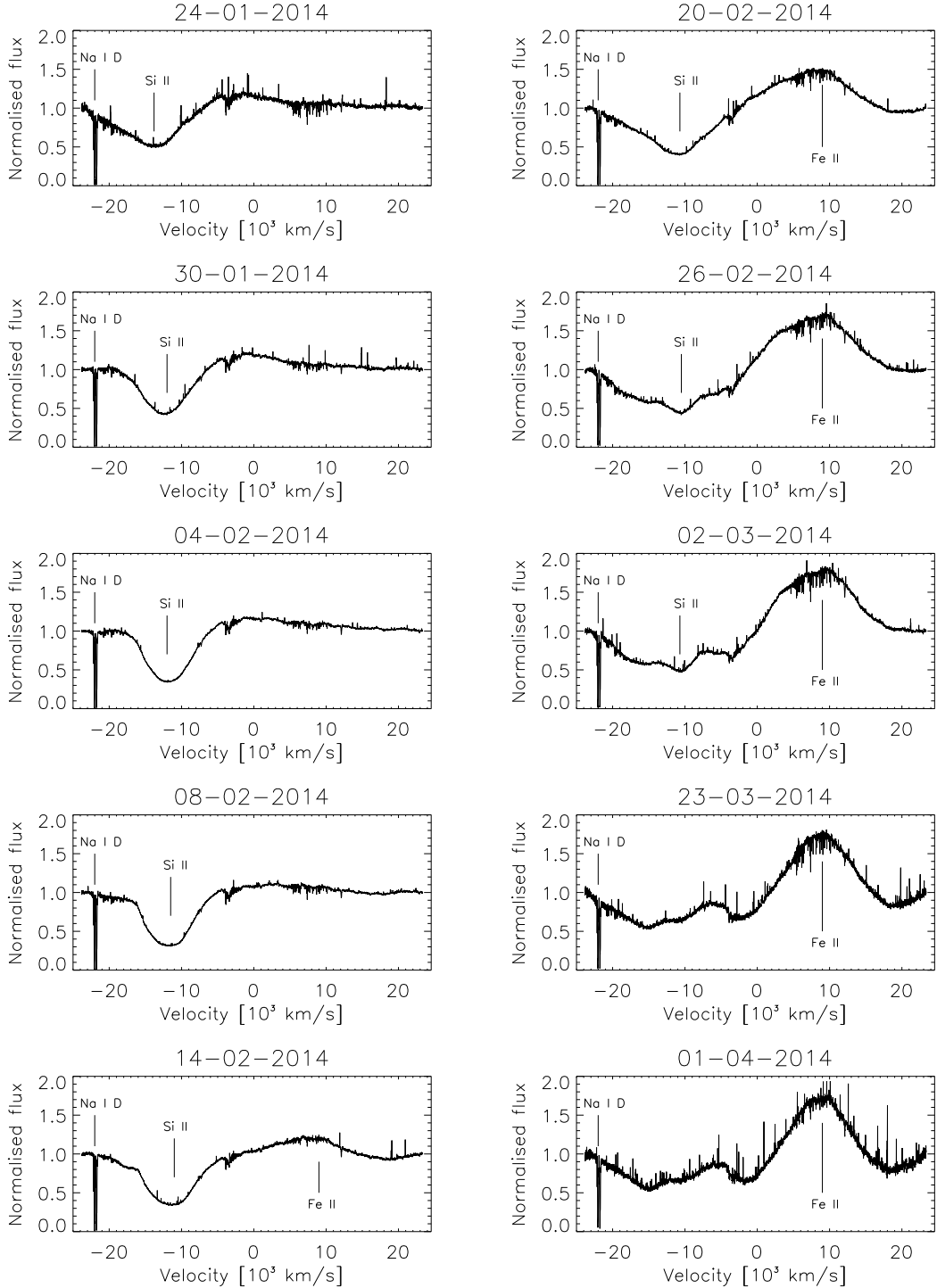}
\caption{The Si II line of 10 observations are shown. These 10 spectra are shown the 
chronology evolution of the time series Si II line at 6355 \AA.}
\label{fig:si_ii_time_evol}
\end{center}
\end{figure*}

The broad Si II P-Cygni line profile is shaped by the ejecta of the supernova, 
its dynamics and its opacity evolution, in a most representative way, 
since this is the strongest feature observed in the SN Ia spectra at
6355~\AA\ during maximum light. 
 
A time series of the wavelength range of this Si II feature is shown in
Fig.~\ref{fig:si_ii_time_evol}. A set of 10 spectra was selected, which best 
represent the evolution and its chronology here. All these spectra are shown 
in their full spectral resolution. They are normalized in flux, and their 
wavelength scales were transformed onto a barycentric 
velocity scale (with respect to 6355~\AA). 

In the first spectra, the Si II feature exhibits a typical P-Cygni profile, which 
is expected for a rapidly expanding, optically still thick envelope. Hence,
the emission is still much less prominent than the absorption part of the profile.
Furthermore, it is clearly seen from the individual line profiles, how fundamentally
the absorption part changes. At the beginning, this is just one consistent broad 
absorption, as of a typical P Cygni profile. Its minimum is located at an expansion
velocity of around $\approx-14,000$~km/s.
This characteristic expansion velocity then shifts slightly to $\approx-11,000$~km/s
until the day of February 20. 
By February 26, however, the Si~II absorption has changed its nature 
significantly, because a small emission feature appears inside of the Si~II
P-Cygni absorption feature.
By April 1, the Si~II feature has come close to disappearance. 

During the later phase an Fe~II emission feature arises in the wavelength range
of the Si~II feature. During the ongoing expansion the envelope becomes thinner
and allows that deeper parts of it shape the spectra. In this way,
the iron peak elements from the inner envelope of the expanding SN Ia envelope
become visible. By the day of
February 26 a small emission feature at around $\approx-7000$~km/s 
has appeared in the absorption trough of
the P-Cygni profile of the Si~II feature.
The emission feature to the right at $\approx+9000$~km/s
is an Fe~II feature, which becomes stronger until April 1. By that time 
the envelope has become optically thin.

\begin{figure}
\begin{center}
 \resizebox{\hsize}{!}{\includegraphics{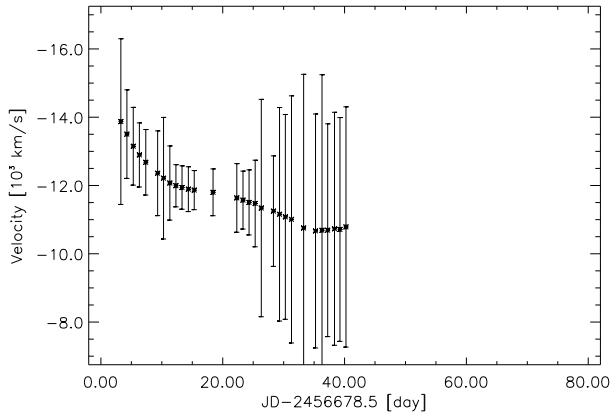}}
\caption{Measured expansion velocities of the Si~II feature at 6000~\AA.}
\label{fig:siiirv}
\end{center}
\end{figure}

\begin{figure}
\begin{center}
 \resizebox{\hsize}{!}{\includegraphics{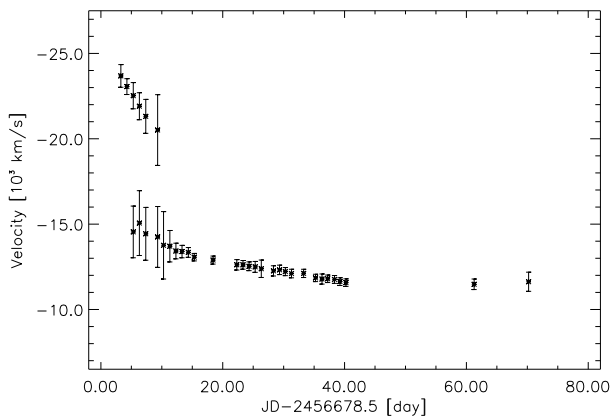}}
\caption{Measured expansion velocities of the Ca~II feature at 8500~\AA.
During the early phase it shows a high velocity component.}
\label{fig:caiirv}
\end{center}
\end{figure}

\subsection{Expansion velocities of Si~II and Ca~II}

Since we obtained a well resolved time series of SN~2014J spectra,
we were able to study the expanding envelope in some detail.
For that, we measured the evolution of the expansion velocities of the Si~II
feature at around 6300~\AA\ and the Ca~II IR triplet feature at around 8500~\AA.

Figure \ref{fig:siiirv} shows the change in the expansion velocity of the
prominent Si~II feature. Starting from a high expansion velocity of 14,000~km/s
on January 24, the expansion velocites decrease while the envelope is expanding. 
Hence, the deeper parts are revealed in the spectra,
which have slower expansion velocities.
When the Si~II is about to disappear, the expansion velocity is about 12,000~km/s.
The photosphere has then passed the layers with abundant silicon.

In Fig. \ref{fig:caiirv}, we show the measured expansion velocities of the Ca~II
feature at around 8500~\AA. Thanks to our early observations, we could study the evolution
of a high velocity component of the Ca~II feature. Its expansion velocity decreases
from 24,000~km/s in our first spectrum down to about 20,000~km/s in the spectrum 6 days
later. 
In their study of near infrared spectra of SN 2014J, \citet{marion15} found 
a similar behaviour for the Ca~II IR triplet expansion velocities.
They measured values for the high expansion velocity feature between 26,000~km/s and 20,000~km/s.
Like in our study, they did not detect a clear high velocity component of the Si~II feature.
All these observations are consistent with the study
of high velocity components in other type Ia supernova performed by \citet{silverman15}.
Observing a high velocity component of Ca~II is much more common than one of Si~II.

The main component of the Ca~II feature starts with an expansion velocity of roughly 14,500~km/s
and decreases to a value of 12,000~km.
As expected in SNe~Ia, the observed expansion velocities of the Ca~II feature
are similar to those measured for the Si~II feature.

\section{The interstellar absorption components of the sodium D-lines}\label{sec:NaD}

%
\begin{figure}
\begin{center}
 \resizebox{\hsize}{!}{\includegraphics[angle=90]{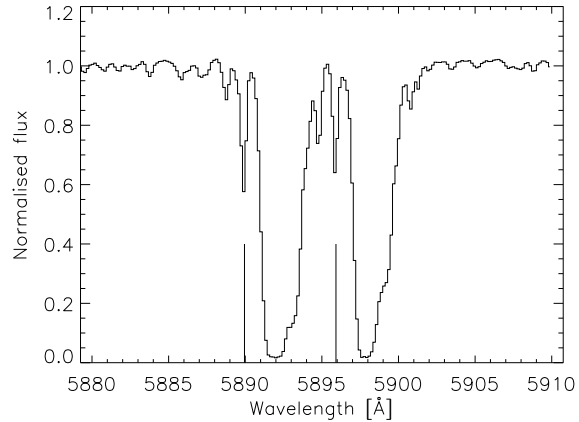}}
\caption{Full resolution capture of the complex IS absorption in the sodium D-line
doublet, from the superposition of the 32 best spectra.}
\label{fig:NaD}
\end{center}
\end{figure}

As can be seen in Fig.~\ref{fig:NaD}, the D-line doublet of Na~I at 5889.95~{\AA} 
and 5895.92~{\AA} shows a complex, multi-component absorption line structure due 
to distinctly different contributions. The telluric absorption lines were not
removed in the presented spectra.
We here study and discuss this very valuable information and offer an 
interpretation in terms of different origins. Within the noise of individual 
spectra, the D-line absorption features appeared invariable over the two months 
or so we succeeded with the spectroscopic monitoring. To minimize noise, 
we superimposed the 32 best SN spectra in the vicinity of 
the Na D doublet (see Fig.~\ref{fig:NaD}), after a barycentric correction was 
applied to the laboratory wavelength scale of each spectrum.

There are essentially four absorption components or groups of even narrower 
absorption features, as have also been seen by other groups in even higher spectral 
resolution (see \citet{graham14,welty14,ritchey15}): 
(a) is a small absorption 
(the right-most feature in Fig.~\ref{fig:NaD}) at a radial velocity shift of 
+245 km/s (barycentric), which is clearly seen in the longer-wavelength Na~I 
D-line but is blended (see below) in the left one. (b) is a very broad and 
saturated absorption trough, prominent in both D-lines. Its distinct, asymmetric 
profile in both D-lines suggests an interesting velocity and density structure, 
which reaches from a sharp edge at +50~km/s to beyond +200~km/s, close to component 
(a). Component (c) is a sharp, medium-strong component well visible in both 
D-lines at $-5$~km/s, while (d) is a 4 times weaker (by equivalent width) component, 
equally sharp and distinct, at $-60$~km/s. It is well visible only for the 
shorter-wavelength D-line, while its longer-wavelength counterpart (d') 
falls into component (a) of the shorter-wavelength D-line.

The uncertainties of the equivalent widths, as measured in the integrated
spectrum are about 0.015 {\AA} for the weak lines 
(components a, c and d, see below) and 0.05 to 0.1 \AA\ for the broad component b.
Individual spectra yielded uncertainties about 6 times as large. Consequently, 
the weak components were well resolved but much affected by noise. Any 
hidden temporal changes, if any, would have had to be smaller than about 25\%
of the absorption of the weak components.

\subsection{Components (c) and (d): distinct ISM clouds}

For component (c) the obvious absorber candidate is galactic ISM. The line 
of sight to SN 2014J probes high galactic latitudes away from the galactic centre, 
in galactic coordinates towards $l=141.4^\circ$ and $b=40.6^\circ$. Around this 
galactic longitude, \citet{mohan04}
observed radial velocities of cold ISM clouds between $-12$ and $+10$~km/s (see their 
Fig.~2), with respect to the local standard of rest (LSR). A small correction 
of +6 km/s applies to translate these veliocities into the heliocentric system 
quoted above. Still, component (c) fits well in here.

Despite its modest appearance, component (c) suffers from some degree 
of saturation, because the two versions of it do not show equivalent widths in 
a ratio of 1:2 (as of the respective f-values) but differ by only a factor of 1.25 
(0.20 and 0.16~\AA, $\pm 0.01$~\AA, approximately). Furthermore, their line widths
are very close to the instrumental profile, which suggests a
very small broadening velocity much under 10 km/s. Therefore, saturation can set in
already with small equivalent widths. 

The analysis of pure absorption lines was formerly applied by us to special problems 
in the field of stellar atmospheres. See \citet{schroeder94} and references given 
there for a brief summary of the simple theory. Given well-known transition 
probabilities (for the Na I D-line doublet we use the $f$ and $g$ values given by
\citet{wiese69}), the equivalent width of a weak, unsaturated line is a linear 
function of only the column number density $N$ of the absorber. However, once saturation
sets in, at least two lines of the same multiplet and with much different transition
probability $f$ are required to solve for both, $N$ and the line-broadening velocity. 
We here use the same simple computer code as \citet{schroeder94}, to calculate the 
profiles of pure absorption lines and their equivalent widths for any such 
parameter pair.

In the case of the two Na~I D (c) components, we find a best-matching solution for 
$N_{\rm Na} = 3.3 (\pm 0.7) \times 10^{12}\;{\rm cm^{-2}}$ and $v_{\rm disp} = 3.5 
(\pm 0.2)$ km/s. This small velocity dispersion matches the lower end of what 
\citet{wilson11} (see their Table 1) find for whole galaxy disks, and presumably 
we indeed see just a single cloud in our own Milky Way. 
We should note, however, that a possibly smaller velocity dispersion
naturally allows for a larger column density due to saturation effects. In fact, 
\citet{ritchey15} find $9.5\times 10^{12}\;{\rm cm^{-2}}$ at 2.0~km/s for this component.
The sodium column density only gives us a lower limit for neutral hydrogen 
of $N_{\rm HI} = 2.1 (\pm 0.5) \times 10^{18}\;{\rm cm^{-2}}$, when using a relative 
sodium abundance of $10^{-5.8}$, since IS sodium is much more prone to ionization 
(much lower ionization potential) than hydrogen. The true value may be almost two 
orders of magnitude higher (see \citet{ritchey15}).

Apparently, the distinct, small and sharp component (d) is of a similar nature
(single cloud) as component (c). Even with the same small broadening velocity, 
it would not suffer much from saturation, as the equivalent width is only 0.05~\AA. 
Accordingly, for its column density we obtain 
$N_{\rm Na} \approx 3 \times 10^{11}\;{\rm cm^{-2}}$. 
At $-60$~km/s barycentric radial velocity, it may appear not very likely to 
be of a galactic origin, but the Leiden/Argentine/Bonn (LAB) H 21cm Survey does show
galactic IS hydrogen here (see \citet{ritchey15}).

\subsection{The signatures of cool ISM in the M82 disk}

Now, can the remaining D-line absorption components (a) and (b) be attributed to 
the M82-internal ISM? And is the observed D-line absorption consistent with the 
IS extinction in the line of sight towards SN 2014J, which mainly arises in 
the disk of M82 (as the line of sight stays clear of most of the galactic IS 
absorption)? The maximum magnitude of SN 2014J of V = 10.5 mag suggests 
$A_{\rm V} \approx 1.9$ mag (with ${\rm M-m} = 27.8$~mag at 3.6 Mpc distance and 
M$_{\rm V, max} = -19.2$~mag for a typical SN type Ia). 
The large reddening (see above) seems to suggest even a larger IS absorption.
 
Using the relation given  by \citet{guver09}, i.e. $N_{\rm H}/({\rm cm^{-2}}) = 
2.2 \times 10^{21} A_{\rm V}/$(mag), a neutral hydrogen column density of $N_{\rm H}$ 
of about $4 \times 10^{21}\;{\rm cm^{-2}}$ should be expected. 
The respective sodium column density would be limited to $N_{\rm Na} < 
6 \times 10^{15}\;{\rm cm^{-2}}$. Due to ionization, the true value could be 
nearly two orders of magnitude lower.

The (here clearly not given) case of an optically thin absorption translates the 
equivalent width of all D-line absorption, nearly 4 {\AA}, into a lower limit for 
$N_{\rm Na}$ of only $ \approx 2 \times 10^{13}\;{\rm cm^{-2}}$. 
Hence, this is indeed fully consistent with an IS absorption of about 
2 mag (or more). Nevertheless, from a profile analysis of a much better 
spectral resolution, \citet{ritchey15} find only $1.8\times 10^{14}\;$cm$^{-2}$.

M82 as a whole has a radial velocity of $+203$~km/s \citep{Chynoweth08}. With respect 
to any Na D absorption components, however, only ISM, which lies in front of SN 2014J, 
becomes visible. SN 2014J is located in the western disk-half, 58'' away from the 
galaxy nucleus. Here, the ISM rotates towards us with (in the rest-frame of M82) 
$v_{\rm rad} \approx$ $-100$ to $-150$~km/s (see \citet{mayya09}). For this reason, 
the D-line absorption should appear 
between (barycentric) radial velocities of +50~km/s to +100~km/s. Remarkably,  
the broad component (b) features a sharp blue edge at +50~km/s, but then reaches out 
with a red wing to component (a) at +245~km/s (barycentric). That material is clearly 
not participating in the general disk rotation but (in the rest-frame of M82) moves 
backwards, towards the SN~2014J.

An answer to such diverse dynamics may be given by the hot superbubble discovered by 
\citet{nielsen14}. Its X-ray emission appears to coincide with the site of SN 2014J. 
The bubble has been inflated to about 200~pc in diameter by the action (hot winds 
and SNe) of some very massive stars formed there perhaps 50~Myrs ago. We suggest that 
the fully saturated part of component (b) between $+50$ and about +140~km/s 
(barycentric) traces the swept-up, cold IS gas in front of that superbubble. 
In the rest-frame of M82, it moves with about $-150$~km/s to $-60$~km/s. Part of 
this gas is pushed towards us by the still 
expanding superbubble, in addition to be rotating towards us with the disk of M82. 
In particular, this would explain the sharp blue edge of component (b): this is the 
cool material closest to the bubble-shockfront approaching us.

The backward moving ISM found in the red wing of component (b) and in component (a)
could be explained, if SN 2014J was not located inside the superbubble, 
but instead a bit {\it behind} it, placing the superbubble at +150~km/s 
(barycentric), that is, in co-ration with the disk, and have its shockfront 
still expanding at about 100~km/s. In this scenario, SN 2014J would also probe some 
ISM pushed out away from us, towards it and so moving against disk rotation. 
The small but distinct component (a) at about +40~km/s in the rest-frame of 
M82 would mark this cold ISM closest to the far-side bubble-shockfront. However,
the highly complex substructure of the IS absorption resolved by \citet{graham14}
leaves, as intriguing as our proposal is, alternative explanations, e.g. 
by a number of individual structures in the long line of sight. And component (a)
may also match the western H~I streamer found by \citet{yun93}, see \citet{ritchey15}.

\subsubsection{Other IS absorption signatures: K~I, Ca~II and DIBs}

\begin{figure}
\begin{center}
 \resizebox{\hsize}{!}{\includegraphics{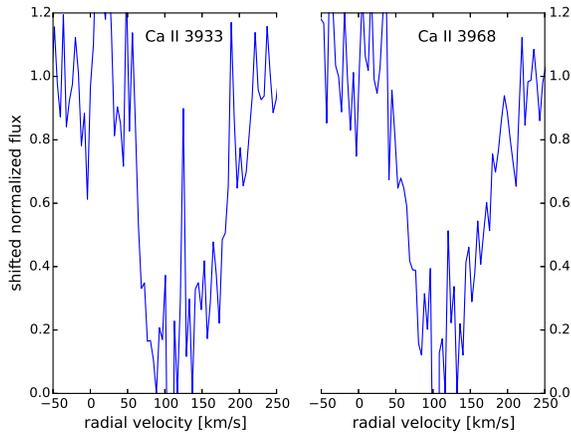}}
\caption{Detection of absorption by ISM for the Ca II 3933 and Ca II 3968
lines in the high resolution spectra of SN 2014J.}
\label{fig:Ca_II}
\end{center}
\end{figure}

\begin{figure}
\begin{center}
 \resizebox{\hsize}{!}{\includegraphics{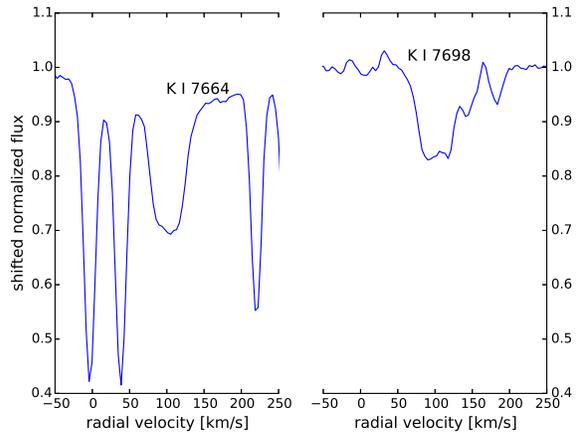}}
\caption{Detection of absorption by ISM for the K I and K I 7698
lines in the high resolution spectra of SN 2014J. The three narrow lines
at the left hand side are telluric lines.}
\label{fig:K_I}
\end{center}
\end{figure}

Interstellar absorption, mostly of component (b) belonging 
to the galaxy M82, is visible in several other lines as well, 
foremost in the Ca II doublet.
Figure \ref{fig:Ca_II} shows two absorption features corresponding
to Ca~II H and K at wavelengths of 3933.66~\AA\ and 3968.5~\AA.
Since these lines are at the lower end of the blue channel,
the spectra at that wavelength range have low S/N.
Obviously, one cannot do a very detailed study.
However, both absorption features are broad, ranging between +50 and +200~km/s,
much like component (b) of the D-lines of the Na I doublet.
We were also able to detected two ISM absorption feature
of K~I at 7698.96~\AA\ and at 7664.9~\AA\ as shown in Fig.~\ref{fig:K_I}.
Note that telluric absorption lines have not been removed from the 
observed spectra.
These feature also show the broad absorption at around +100~km/s
corresponding to the broad IS absorption feature of the sodium D-line doublet.
This broad feature is due to absorption of ISM in M82.
We were also able to detected CH+ at a wavelength of 4232~\AA.

The high resolution of our spectra allows us to check even for  
unidentified interstellar absorption features, the so-called Diffuse Interstellar 
Bands (DIB). So far, the responsible elements or molecules have not been 
identified, which is a motivation for further high-quality observations of DIBs.
We were able to detect a few of them in our best high resolution spectra of the 
supernova SN~2014J.

\begin{figure}
\begin{center}
 \resizebox{\hsize}{!}{\includegraphics{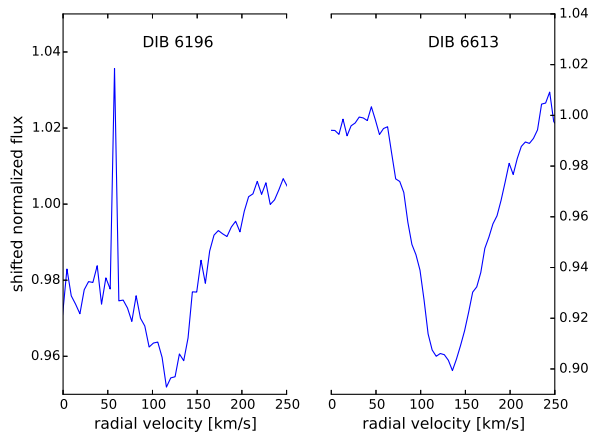}}
\caption{Two features of DIBs observed in the high resolution spectra of SN 2014J.
The left hand side shows the feature at 6196~\AA, and on the right hand side 
the DIB 6613 \AA.}
\label{fig:dib}
\end{center}
\end{figure}
\begin{figure}
\begin{center}
 \resizebox{\hsize}{!}{\includegraphics{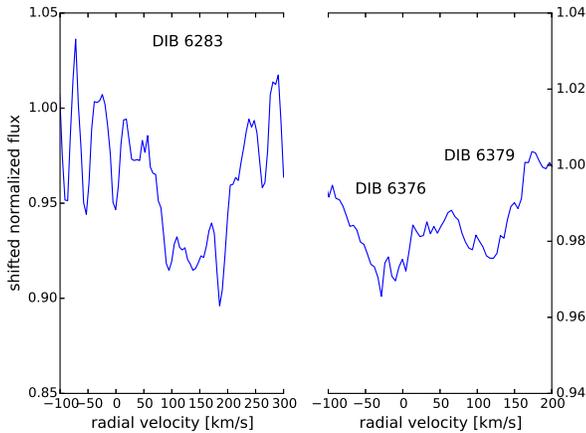}}
\caption{Three more features of DIBs observed in the high resolution spectra of 
SN 2014J. We found DIB features at 6283~\AA, as shown on the left hand side, 
and on the right hand side two features at 6376 and 6379~\AA.}
\label{fig:dib2}
\end{center}
\end{figure}

Since the DIB features are present in all our spectra, we again added them up to 
obtain just one final spectrum with further improved S/N, As shown in Fig. 
\ref{fig:dib}, we find two DIBs at wavelengths of 6196 \AA\ and 6613 \AA. The
DIB wavelengths have been taken from the study of \citet{hobbs09}.
Again, the plot indicates velocities in the range of roughly +50~km/s 
to +200~km/s, as found in the broad (b) absorption feature of the Na D doublet.

In this combined spectrum of SN~2014J, we also found several other DIBs. 
Figure \ref{fig:dib2} shows the DIB of 6283~\AA\ and the double feature of
DIBs at 6376~\AA\ and 6379~\AA, in the same range of +50~km/s to +200~km/s.
All these feature were also observed by \citet{welty14} and several 
other authors from high 
resolution spectra of SN 2014J. 
We were unable to detect the 5780 and 5797 \AA\ DIBs since
these features unfortunately lie exactly in the wavelength gap of
the HEROS spectrograph.

\begin{table}
\caption{Measured equivalent widths of the ISM and DIB features observed in the
spectra of SN 2014J.}
\centering
\begin{tabular}{c c}
\hline
Feature & Equivalent Width [\AA]\\
\hline 
DIB 6283 & 0.27 $\pm$ 0.04\\
DIB 6376 & 0.065 $\pm$ 0.013\\
DIB 6379 & 0.037 $\pm$ 0.007\\
DIB 6613 & 0.2 $\pm$ 0.01\\
K I 7664 & 0.35 $\pm$ 0.05\\
K I 7698 & 0.28 $\pm$ 0.03\\
\hline
\end{tabular}

\label{table:ew}
\end{table}

In Table \ref{table:ew}, we present the measured equivalent widths of the ISM 
and DIB features observed in our spectra of SN~2014J. The spectra in the
wavelength range of the Ca~II H \& K absorption lines are too noisy
to able to obtain reasonable equivalent widths.
Our measured values for the equivalent widths of the DIBs 
are in the same range as found by \citet{welty14} and \citet{graham14}.
However, for the DIB~6283~\AA, the strongest of the observed DIBs in our spectra,
\citet{welty14} found a much higher value for its equivalent width, while
\citet{graham14} found a value comparable to our. The reason might be
the broad wings of this line as seen in the spectra of \citet{welty14}.
In our spectra, the telluric lines make it difficult to measure equivalent
widths including the broad wings (see Fig.~\ref{fig:dib2}). 
The IS absorption in both lines of K~I are stronger than the DIBs.

\section{Conclusions and discussion}

We presented a time series of 33 high resolution spectra of the
Type Ia supernova 2014J in M82 observed with the TIGRE
telescope. We obtained these spectra with the HEROS
\'echelle spectrograph, which covers the wavelength range from
3800~\AA\ to 8800~\AA, with only a small gap at around 5800~\AA. 
The long monitoring period of about two months, well including
maximum light of SN~2014J, in combination with the high quality of 
the spectra, enables us to empirically describe the 
spectral evolution of this supernova Ia in detail.
By its spectral features and their behaviour SN~2014J is 
very typical for a Type Ia SN.

During maximum light, red-channel SN~2014J spectra show broad P-Cygni profiles 
caused by the strongest lines of Si~II and Ca~II. In the blue channel
we observed the typical features of S~II, Mg~II, Fe~II and Ca~II.
We studied the evolution of the prominent P-Cygni profile features of Si~II
at around 6300~\AA\ in some detail.
It shows a decrease in the expansion velocity in time as it is expected from
rapidly expanding SN envelopes.
The first spectrum from January 24 shows a minimum in the absorption trough at
an expansion velocity of $\approx -14,000$~km/s. The expansion velocity
decreases, and on February 20 it has dropped
to a value of $\approx -12,000$~km/s.

The Ca~II infrared triplet feature at about
8500 \AA\ shows a high expansion velocity component during the first days of observation
with expansion velocities of 24,000~km/s measured on January 24 decreasing to 20,000~km/s
right before this component disappears.
The main expansion velocity component shows a similar behaviour as the Si~II feature.
The expansion velocities decreases from an inicial value of 14,500~km/s down
to 11,0000~km/s during the later phases of the expansion of the envelope.

Due to the significant amount of interstellar absorption and reddening 
of SN~2014J, the later blue-channel spectra  suffer from a quite low S/N.
In the well-exposed red channel, however, those later spectra document 
the transition from the optically thick to the thin phase of the expanding envelope
very well. Now, emission features are dominated by Fe~II lines from deeper layers.
At the same time, in the red channel, the Si II feature disappears, while a 
Fe II feature arises in the respective wavelength region. 

The reason for the rise of these Fe II lines lies in the layered structure of 
the SN debris, combined with a decreasing opacity: the quasi-photosphere moves 
inwards and passes from the outer layers of intermediate mass elements like
silicon and sulfur 
to deeper layers with iron peak elements, stemming from the inner envelope of the 
exploded progenitor White Dwarf. In this way 
our dense time series allows us to basically \emph{scan} the abundance properties 
of the entire SN envelope, especially its abundances structure. Achieving this 
requires, of course, a huge amount of modeling work. Every spectrum of the series needs to be 
analyzed and/or modeled with a suitable (1D or even 3D, non-LTE, dynamic) atmosphere 
code. In a future work, using the \phx code \citep{hauschildt92,hbjcam99},
we will do this work for the whole time-series of spectra, in oder to obtain
a detailed and quantitative abundance structure of the envelope of SN~2014J as
a case study for a SN of Type Ia. 
 Since the different explosion models predict slightly different abundances and also
a differently layered abundance structure, this detailed abundance determination might exclude or favor
one or some of the suggested explosion mechanism for a supernova of Type Ia.
Some of the here presented spectra have already been 
used to identify the feature, which causes the secondary maximum in the $I$-band
light curve of Type Ia supernovae \citep{jack15}. 

Furthermore, the high resolution of $R\approx 20,000$ of the obtained spectra 
allows for a study of the sharp ISM absorption features in the SN~2014J spectra. 
The Na~I D doublet shows different substructures in the ISM of our galaxy, 
as well as of the host galaxy M82. There is a small feature of ISM of M82 at a
high velocity of about +245~km/s (barycentric). A very broad absorption feature 
at velocities of +50~km/s to +200~km/s is apparently caused by ISM in M82, perhaps
even related to the expanding environment around a superbubble just in front of the SN.
We also found two sharp absorption features at velocities of about $-5$~km/s and 
$-60$~km/s, both are of a galactic origin. 
We are also able to identify the same broad ISM absorption of M82 between 
+50~km/s and +200~km/s in the principal lines of Ca~II H \& K at 3934 and 3968~\AA\
and K~I at 7664 and 7699~\AA.

Thanks to the high quality of the spectra, we also identified a few DIBs in 
the spectra of SN~2014J, confirming other groups results. 
We found signatures of these features at 
wavelengths of 6196, 6283, 6376, 6379 and 6613 \AA\ and measured 
the equivalent widths of these features. Hence, we
can fully confirm the observations made by \citet{welty14} in their
analysis of high resolution SN~2014J spectra. The radial velocities
of the observed DIBs again are fully consistent with the broad Na D absorption
feature of M82 in the range of velocities of +50~km/s to +200~km/s.

\section*{Acknowledgments}

Our collaboration and work was much helped by travel money from 
bilateral (Conacyt-DFG) project grant No. 192334, as well as by 
Conacyt mobility grant No. 207662. Furthermore, we are very grateful for the
technical support of TIGRE (infrastructure, Internet connection, maintenance 
of hardware and software) by, namely, the DA-UG engineers Filiberto Gonz\'alez 
and Ir\'an Montes, as well as by UG's general technical support (namely by 
colleagues at DSTI and Infraestructura). We would also like to thank the anonymous
referee for many very helpful comments and suggestions.

\bibliographystyle{mn2e}
\bibliography{all}

\label{lastpage}

\end{document}